# Robust Watermarking using Diffusion of Logo into Autoencoder Feature Maps

Maedeh Jamali[1], Nader Karim[1], Pejman Khadivi[2], Shahram Shirani[3], Shadrokh Samavi[1,3]

[1]Electrical and Computer Engineering, Isfahan University of Technology, Iran,
[2]Computer Science Department, Seattle University, Seattle, USA,
[3] Electrical and Computer Engineering, McMaster University, Hamilton, Canada

*Abstract*— Digital contents have grown dramatically in recent years, leading to increased attention to copyright. Image watermarking has been considered one of the most popular methods for copyright protection. With the recent advancements in the application of deep neural networks in image processing, these networks have also been used in image watermarking. Robustness and imperceptibility are two challenging features of watermarking methods that the trade-off between them should be satisfied. In this paper, we propose to use an end-to-end network for watermarking. We use a convolutional neural network (CNN) to control the embedding strength based on the images' content. Dynamic embedding helps the network to have the lowest effect on the visual quality of the watermarked image. Different image processing attacks are simulated as a network layer to improve the robustness of the model. Our method is a blind watermarking approach that replicates the watermark string to create a matrix of the same size as the input image. Instead of diffusing the watermark data into the input image, we inject the data into the feature space and force the network to do this in regions that increase the robustness against various attacks. Experimental results show the superiority of the proposed method in terms of imperceptibility and robustness compared to the state-of-the-art algorithms.

## 1. INTRODUCTION

Due to easy access to the Internet during the past decades, sharing digital content has become a prevalent practice. Contrary to analog versions, digital media can be copied with the same quality as the original one. Image editing tools can threaten digital media's integrity. Digital tampering is now an easy task. Unauthorized copying and distribution of images are on the rise. Hence, the need for copyright protection has been emphasized in the literature [1]. One of the most reliable and popular solutions to address these security issues is digital watermarking. Digital watermarking was introduced in 1979 for anti-faked purposes [2], recognizing original documents from the fake ones. In watermarking methods, the ownership of copyrighted media is identified by embedding invisible information into an image extracted later [3]. Besides the ownership authentication, watermarking has other applications such as concealing patient's information in their medical images for medical procedure matching and identification [4], broadcast monitoring [5], device control [6], and copy control purposes [7].

Usually, the hidden information on watermarked images is sensitive and vulnerable to image processing transformation and enhancement like image compression, cropping, or undesired artifacts such as transmission noises. Hence, three key attributes are considered for evaluating the watermarking algorithms: imperceptibility, robustness, and capacity [8]. These features act contrary to each other, and the efficiency of watermarking methods highly depends on the trade-off between them. Robustness means the preservation of hidden information against image processing attacks. On the other hand, imperceptibility aims to make the watermarked image indistinguishable from the original image. The amount of hidden information (capacity) is also essential in some applications, such as medical procedures. Different classifications can be considered for watermarking methods based on the embedding domain, the embedding method, and the extraction procedure. Concerning the extraction process, watermark methods will be classified into three classes, blind, semi-blind, and non-blind [9-10]. Blind methods do not need the cover image during the watermark extraction process, while the two other groups require extra side-information alongside the extraction phase. Therefore, blind methods are practically preferred to semi or non-blind methods, although blindness will increase their complexity and may affect their imperceptibility or robustness.

The type of embedding domain is another categorization for watermarking methods. Two categories of spatial domain and frequency domains are usually considered. In spatial domain algorithms [11-14], data are directly hidden in the image pixels. However, these methods are generally not robust against typical image processing attacks. In contrast to the spatial methods, frequency-domain algorithms [15-33] are more robust against different attacks [8].

Recently, machine learning algorithms have been used in watermarking applications, as they provide efficient solutions for embedding and extraction of the watermarks [34-37]. These methods use learning tools for a specific part of watermarking, such as attack detection [34], predicting neighbor pixels relationship for watermark extraction [35], parameter optimization [36], and frequency domain coefficients prediction [37]. One of the most popular machine learning tools is Convolutional Neural Networks (CNN), which have attained considerable attention in various computer vision applications such as object detection [38], image classification [39], and pattern recognition [40]. Using CNN in watermarking is new, and recently, some works used deep networks in watermarking [41-46].

In this paper, we introduce end-to-end trained blind watermarking using CNN encoder-decoder framework. The proposed framework can prepare the watermark data based on the input image's content. Hence, it has a considerable capability of maintaining imperceptibility and robustness in comparison to state-of-the-art methods. The proposed framework consists of two parallel networks: preparing a watermark alongside the main network for embedding and making a watermarked image. It also has an attack layer that tries to simulate image processing attacks. Using the attack layer as a part of the learning system helps the framework increase the robustness of the watermarking method. This improved robustness is achieved since attacks are considered during the training phase.

Traditional watermarking methods use fixed algorithms in the transform domain, such as changing or swapping coefficients. However, in the proposed method, we embed the data in feature maps. Based on the input image and the type of attacks, the proposed method learns to use different approaches such that it has had the highest efficiency based on the definition of the loss function. The rationale for selecting CNN for watermarking is the ability of the network to represent hidden data such as weight kernels, which makes it almost impossible to retrieve the information due to unknown hyperparameters of the network. Therefore, the watermark will only be known to the sender and receiver. This feature makes our method highly robust against hackers who want to change the data. Furthermore, the proposed CNN-based network accepts primary watermark data and prepares it based on the input image to maximize imperceptibility. Hence, the watermark remains robust against various attacks. The proposed framework is a multi-objective network where the loss function tries to create a trade-off between the objectives with opposite aims. Moreover, in our method, the network diffuses the watermark over the entire image. The watermark data will be diffused inside the whole image by replicating the input watermark (with the same size as the input image). Hence, different parts of the image will have the information of the watermark, and the watermarked image attains its robustness against various kinds of heavy attacks. Although significant parts of the image are deleted or corrupted in some attacks, the network could still retrieve the hidden data.

To summarize, the major contributions of the proposed method are as follows:
1. Using a network that is end-to-end trained for embedding, manipulating, and extracting the watermark.
2. Using a parallel network to prepare the watermark based on the input image's content to maximize the imperceptibility.
3. Injecting the watermark into feature space to have better robustness against attacks
4. Considering global information besides local information to have robust watermarking using dilation layer that contributes more pixels information in embedding.

The rest of the paper is organized as follows: Section 2 is dedicated to the literature review. Section 3 introduces the details of the proposed method. The experimental results are shown in Section 4. The paper is concluded in Section 5.

## 2. LITERATURE REVIEW

Since the advent of watermarking, there has been a significant amount of research to improve the main characteristics, such as fidelity, secrecy, capacity, imperceptibility, and robustness. At first, the spatial domain is used for embedding, and the information is embedded in the cover image by manipulating the pixel image [11-14]. Substitution of LSB and the randomized process is used in [11]. They developed a genetic algorithm to find the rightmost LSBs of the cover image. For improving imperceptibility, the entropy of blocks and histogram is used in [12]. First, they divide the cover image into blocks, and some blocks are collected based on their entropy. Then embedding is performed using the histogram shape methods. In [13], a Gaussian low-pass filter is applied as a preprocessing in the embedding phase. Several gray levels are randomly removed using a secret key, then, using the selected gray levels, a histogram of the filtered image is made. Choosing pixel groups with the highest number of pixels based on the built histogram is the novelty of this method. They consider a safe band between selected and not selected pixels, and the chosen pixels are candidates for embedding. The proposed algorithm in [14] is a hash-based method that is robust against alternation such as rotation and cropping. The keynote of this method is that if the tampering is sufficiently sparse, it can be localized using convex optimization problem-solving.

Another group of watermarking methods uses the transform domain that embeds information in the frequency domain by manipulating frequency coefficients. They have better performance than the previous group and effectively preserve the visual quality of the host image [15]. Many different transform domain have been used such as discrete Furrier Transform (DFT) [15], Discrete Cosine Transform (DCT) [16-21], Discrete Wavelet Transform (DWT) [22-27], Contourlet Transform (CT) [28-31] and Hadamard transform [32-33]. The authors of [15] have used DFT to manage global geometric distortion. In [16], the image is divided into $8 \times 8$ blocks, and the DCT of blocks is calculated before modifying the DC coefficients. They use Arnold transform in addition to chaotic encryption to add double-layer security to the watermark. Parah *et al.* [17] introduce a DCT based algorithm by partitioning the input image into $8 \times 8$ blocks, then transform each block into DCT and select some blocks using a Gaussian network classifier. Eventually, the DCT coefficients are changed based on watermark bits. In [18], a DCT-SVD based algorithm is introduced. A luminance mask of the cover image is made in the first stage. In the embedding phase, they modify the singular values of the DCT of the original image with singular values of the produced mask. They also use a genetic algorithm for finding the control parameters. Tian *et al.* [19] use the combination of CT, DCT, and SVD for embedding. First, the input image is decomposed by one-level CT, and its low-frequency sub-band is divided into eight by eight non-overlapping blocks. DCT transforms each block, and some middle-frequency DCT coefficients are selected to make the carrier matrix. Eventually, the watermark is embedded by modifying the largest singular values of two carrier matrices.

Some methods are proposed in the DCT domain using fuzzy systems [20-21]. The authors of [20] use a neuro-fuzzy model for embedding and extraction. They inject the human visual system (HVS) parameters into the fuzzy system and utilize the output of the fuzzy-neural framework as the strength-factor for embedding. Jagdeesh *et al.* [21] propose fuzzy inference models for calculating the strength factors using HVS.

An enormous number of algorithms have used DWT as a transform domain for watermarking, and the middle and high-frequency coefficients are generally considered for embedding [22]. A DWT-based algorithm is used in [23]. They add pseudo-random codes to large coefficients in a middle and high-frequency band. A quantized-based algorithm is proposed in [24]. The angel quantization of gradient vectors in different wavelet scales is considered for embedding. A geometrical model is proposed in [25] for embedding. They extract eight samples of wavelet approximation coefficients from each block of the image and built two line segments in a two-dimensional space. The authors in [26] propose a combination of DCT, DWT, and fuzzy systems for embedding. They first transformed the image into DWT for two levels and then divide the transformed image into $8 \times 8$ blocks and calculate the DCT of them. Three attributes related to HVS are fed to a fuzzy system to calculate each DCT block's strength factor. Then, they manipulate the DCT coefficients based on these adaptive strength factors and watermark bits. Makbol *et al.* [27] used a combination of SVD and DWT for embedding, and singular values are used as secrete keys.

Other types of transform domains that researchers use are CT and Hadamard transform. An adaptive blind method is proposed in [28] in which the DCT coefficients of CT are considered for embedding. They determine an adaptive strength factor for each block using entropy and some other attributes of each block. First, they apply a two-level CT on the input image. They divide the approximate image into blocks in the first level. Using their proposed edge detection method, they extract the important edges. Then, part of the image with high edge concentration is considered as a candidate region for embedding. In [29], a CT-based domain is proposed. They modeled the contourlet coefficients based on a "t-location-scale" distribution. They demonstrate that the "t-location-scale" distribution has high efficiency in modeling the coefficients. The authors of [30] choose the CT domain for embedding. For more robustness, they manipulate the DCT coefficients of CT blocks. In [31], the maximum likelihood method based on normal inverse Gaussian (NIG) distribution is used for extraction. Li *et al.* [32] proposed a watermarking scheme for color images using quaternion Hadamard transform and Schur decomposition. In [33] a Hadamard transform is used for embedding. They analyze the Hadamard's coefficients and find a bit-plane that satisfies transparency and robustness, and then the watermark is redundantly embedded in the selected bit-plane.

Due to the high performance of machine learning methods in watermarking, they have recently become prevalent. A learning-based watermarking algorithm is proposed in [34]. They learn the influence of different attacks on the image. Based on this effect, manipulate different parts of the frequency spectrum that have higher resistance against these attacks. They redundantly embed the watermark in a different part of the frequency domain. Using their attack classification method, they extract the watermark from the part of the frequency domain that is less damaged by the detected attack. A combination of extreme learning machine (ELM), online sequential extreme learning machine (OSELM), and weighted extreme learning machine (WELM) have been used in [35]. They utilize machine learning techniques to learn the relationship between neighbors' pixels to extract watermarks. An optimization-based algorithm using machine learning methods for finding the best embedding strength factor is introduced in [36]. First, a DCT based algorithm is proposed. Then an artificial bee colony is applied on watermarked images as a training process to find the optimum strength factor. A blind learning-based algorithm for classifying the embedded bits is proposed in

[37]. This method tries to adaptively modify the decoding strategy to increase robustness against anticipated attacks. The watermark extraction is performed as a binary classification problem.

Heretofore, previous works have used deep networks for one phase of the detection, such as determining the strength factor for encoding [41], embedding [42], or in the extraction phase [43]. However, in two recent works, they modeled an end-to-end network that used the network for whole watermarking [44-45]. Kandi et al. [42] proposed a non-blind watermarking using CNN autoencoder to hide data in feature maps, using a predefined embedding algorithm. An end-to-end watermarking method is proposed in [43] that embeds data in blocks of the image, similar to traditional methods with uniform local embedding. The authors of [44] present an end-to-end network for steganography and watermarking using the Generative Adversarial Networks (GAN) and CNN. They try to spread the watermark data in all pixels of the input image by replicating the data, and each pixel has all the information about the entire watermark, which can produce some artifacts. However, the GAN helps the framework to prevent these kinds of artifacts from watermarked images. In [45], an end-to-end autoencoder is proposed that simulates a DCT layer in their deep network. After concatenating the input image and the watermark, this combination is used as an input to the DCT layer, and after transforming, the embedding is performed. Due to using the DCT layer, the proposed structure has good robustness against JPEG attack. In contrast to the mentioned works, in our proposed method, we try to prepare the watermark data based on the input image to have the lowest effect on transparency and be robust against image processing attacks. Authors in [46] combine over-complete dictionaries and create redundancy in the signal representation in the transform domain. They proposed a random matching pursuit algorithm that facilitates watermark embedding and extraction.

## 3. PROPOSED WATERMARKING FRAMEWORK

In this section, we explain our proposed method in detail. Our method is an end-to-end CNN that tries to embed, attack, and extract in one framework as a blind watermarking method for grayscale images. It is trained as an end-to-end network to have blind secure watermarking. This network contains two consecutive convolutional neural networks for embedding and extraction. We also have a distinct CNN alongside the main framework for preparing the watermark data, which helps the network improve the extracted image's transparency. Implementing an attack layer to simulate image processing attacks as a network layer guarantees robust watermarking and makes end-to-end training easy. We first explain the whole deep network structure and the details of each part (embedding, attack layer, and extraction) in subsection 3.1. The evaluation metrics and training of the network will be discussed in subsection 3.2.

*3.1. Network Structure*

Figure.1 indicates the block diagram of the proposed method, which includes three main modules: embedding module using CNN, attack layer for simulating image processing attacks, and extraction module for extracting the hidden data. The embedding module accepts two inputs and embeds the watermark in the input image. The extraction module has to extract the watermark data from the reconstructed image after image processing attacks. In the following, each of these three modules will be explained.

*3.1.1. Embedding module*

The first part of the proposed network is responsible for hiding the watermark information into the cover image, but in contrast to traditional methods, which embed information in a transform domain, we have used a CNN to conceal information in feature map space. This would be more robust than typical methods and can be resistant against changing or unauthorized accessing because no one knows the structure of the training network.

The embedding module includes two separate networks, one for preparing the input image and the other one for making the watermark ready based on the cover image's content that adopts adaptive embedding to the cover image. The main part of the embedding module is an auto-encoder that receives the host image as an input. The input will pass from convolutional layers but with the same size as the input. These convolution layers help us extract the input image's main features in each layer, such as edges and other meaningful attributes. Figure. 2 indicates the embedding network that is fed by the input image.

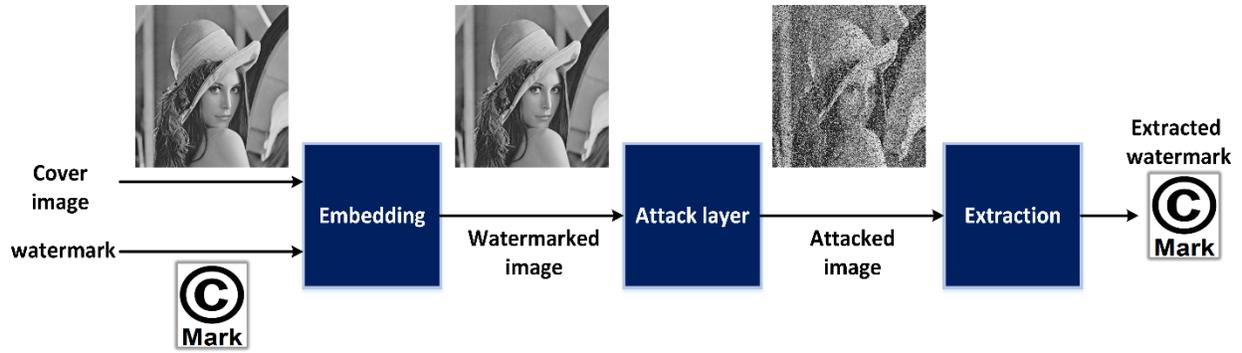

Fig. 1. Block diagram of the proposed CNN-based watermarking method.

As can be seen in Fig. 2, the encoder part contains two distinct networks that link to each other by concatenating the networks' output. The lower network in Fig. 2 is a CNN whose primary responsibility is to prepare the input image for embedding. The network's input is a grayscale image with size $H \times W$ ($H = W = 32$) that is fed to consecutive convolution layers. The input image can have different sizes (dividable by $32 \times 32$) and should be divided into $32 \times 32$ blocks. Hence, we need a reshape part at the first stage of our network. A batch of $32 \times 32$ blocks of the input image is fed into the network. All the input images are normalized into [0, 1]. We do not use pooling layers because we want to keep all information of the layer's input and keep the transparency of the reconstructed image. We have a ternary block [44] in our network's layer that is a combination of convolution layer, followed by activation layer and batch normalization (BN) [47]. Each input to layer will pass from this block. BNs help the network to learn faster, and it also improves accuracy.

We use dilated convolution during learning. Dilated convolution helps the network to aggregate multi-scale contextual information without missing resolution [48]. This attribute helps our network to have a global view of information and expand the hidden data in a larger area. Hence more parts of the feature map will contain the information of the watermark data. Wide distribution is helpful, especially against attacks when some part of data will lose. So other parts of the image that have the information can participate in rebuilding the data.

After suitable numbers of convolution layers, the encoded image (feature map) will be produced by the encoder part in Fig. 2, and it has the same size as the input. The watermark data will be concatenated with the output feature map in this part. This has several advantages. First, the network may memorize the input image due to removing the pooling layer in our network structure, and the accuracy will not be fair. But in this phase, the watermark data is concatenated with the feature map before sending it to the decoder part. This works as a noise (because a different random watermark is used for each training input during learning) that prevents the network from memorizing and forces the network to learn. Second, the hidden data will be fused with feature map data that helps the robustness of the watermarking. Because no one knows about the embedding process and the network structure, thus cannot manipulate it. Third, the proposed framework is an end-to-end network that tries to improve robustness and imperceptibility together, and the loss function controls the trade-off between them. So the network has to change the network's weights and feature maps to maintain transparency and be robust against attacks. Therefore, some parts of weights will have larger values that will better keep the information in different conditions. This helps the whole network to automatically controls all desired features.

After concatenating the encoded image with a watermark, the produced feature maps will be fed to the decoder part that tries to reconstruct the input image. This part has a similar structure as the encoder part, but with one more layer to be able to reconstruct the watermarked image with visual similarity to the input image.

The upper part of Fig. 2 shows a CNN that is fed by watermark data. We name it the watermark preparing network (WPN). This network works in parallel with the lower auto-encoder. It is responsible for preparing the watermark such that it can be extracted better and have lower effects on image visual quality. During the training phase, the network's input is a random watermark string with a size $4 \times 4$ that is smaller than the cover image. The structure of the WPN is similar to the encoder part of the auto-encoder network by ternary blocks and without pooling. The network's output has the same size as the hidden input data. The network has learned to emphasize some parts of the data more than other parts during learning, so we can see that the network's output has a different value than the main hidden data. Considering a network in addition to the main embedding network and preparing the watermark before embedding has positive effects on the robustness and quality of the produced image. A good idea to have a better watermarking algorithm is to force the whole host image to have hidden information. To this aim, we change the size of the WPN's output to $H \times W$ by tiling. By replicating the processed watermark along width and length, we form a $H \times W$ tiled watermark. This can be considered as redundancy too. Next, the tiled watermark map is concatenated with the feature map, and they are fed to the decoder part of the auto-encoder network. During the decoding, all the feature maps and the watermark will be fused into a feature map. Finally, the decoder's output is the watermarked image that contains the watermark information. In this part, we need another reshaping to put the image blocks beside each other and make the new image similar to the input image with the same size. After embedding the watermark, we should add image processing attacks and then extract the watermark. For this aim, we add an attack layer as a part of the network to simulate digital attacks during learning and help the network develop robustness against them. The attack layer is explained in the next section.

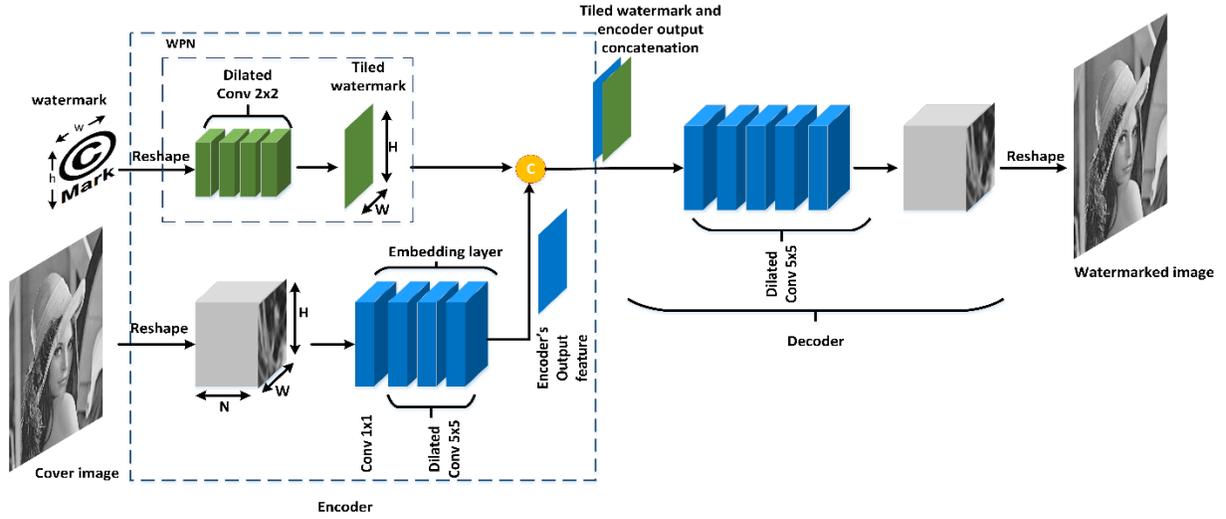

Fig. 2. The embedding module of the proposed method. This network acts as an auto-encoder that the encoder part contains two networks. The lower part is used to prepare the input image for embedding. All feature maps have the same size as the input image. The upper part of the encoder is a CNN that prepares a watermark for embedding based on the image content. The watermark is tiled as the same size as the input image to diffuse the data in all feature maps.

*3.1.2. Attack layer*

In this section, we explain the details of the attack layer that is shown in Fig. 1. Considering the attack layer as a part of the network helps the network learn more complicated watermarking patterns, which are more robust against various attacks. Another advantage of using the attack layer during learning is mitigating overfitting and preventing the network from memorizing. We consider a combination of attacks to have a multi-attack network that is robust against different kinds of attacks. A random selection method is used to select one attack each time and send it during training. Due to its randomization, each attack has the same chance to select. This strategy helps the network see various attacks during learning, so the network tries to find suitable parts of the feature space robust against all using attacks. Eventually, we will have a network with robustness against different attacks. For example, some of them manipulate high-frequency coefficients, and some of them change the low-frequency parts. Hence, we will have a general multi-attack network by selecting the combination of attacks with both features. We briefly explain each attack in the following.

A. *Noise attacks*

Gaussian noise, salt and paper, and uniform noise can be considered as noise attacks in which a random noise will be added to the input image in each iteration of training. However, for salt and paper, we should randomly select a ratio of pixels and change their value to white or black. We used a binomial distribution for selecting and changing the values of pixels using two masks (one for producing white pixels and one for producing black pixels) . The same noise patterns are used for all images in a batch, but they would be different from images in another batch.

B. *Gaussian blurring attack*

We have used a depth-wise convolution layer that applied a filter on each input channel independently for the blurring attack. We have fixed weights for this convolution layer, and these weights can be produced for various kinds of impulse response filters like Gaussian or sharpening filters.

## C. JPEG attack

Most of the mentioned attacks are differentiable and can be easily implemented as a network layer. However, JPEG compression has a quantization procedure that is not differentiable. Hence, for implementing it as a network layer, we need to approximate the quantization. JPEG uses a DCT transform and quantization for compression. First, it divides the input image into $8 \times 8$ blocks, and then each block is transferred into DCT. Then the DCT coefficients will be divided by a quantization matrix. This quantization process is not differentiable and so cannot be used in gradient descent optimization. During quantization, usually, the high-frequency coefficients become lower, and the theoretical concept of quantization also means limiting the amount of information that can be transmitted. Figure. 3 shows a method we used for implementing a JPEG attack. Fig. 3 (a) is an $8 \times 8$ DCT block with different values. Fig. 3 (b) indicates a quantization block sample. The quantization values become higher from left to right and up to down in a quantization block as the color shows in the image. By dividing the DCT coefficient by these quantization values, the high frequencies of the DCT block become lower (near-zero), and we can ignore them. Instead of using this quantization table which is not derivative, we use the third block in Fig. 3 that keeps some high coefficients from the low frequency in the DCT block. Hence, we can simulate the JPEG compression by ignoring some high-frequency coefficients. We implement DCT transform using convolution layer with stride 8 and kernel size 8, and the kernels are the basis vector in DCT transformation for implementing this attack layer. Then we keep a few DCT coefficients from the low-frequency part and make the high-frequency coefficients zero in a zigzag manner. Because we drop many coefficients, our layer can simulate JPEG attacks with low quality during learning.

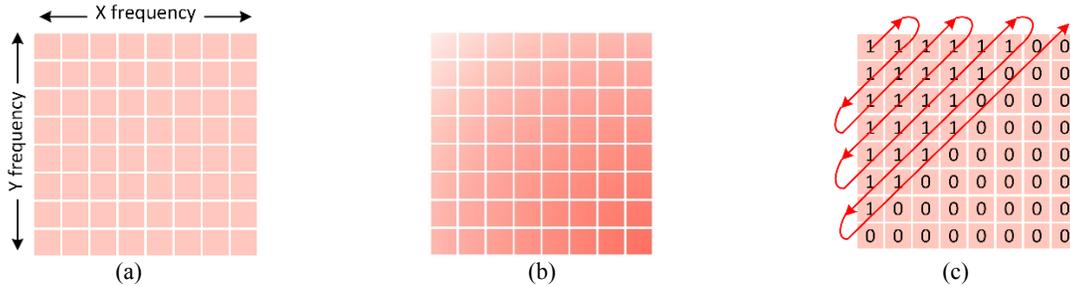

Fig.3. DCT, quantization coefficients values and simulated quantization coefficients for using in a deep network, (a) DCT block, (b) Pseudo-matrix of quantization values that high frequencies are divided with higher values, (c) simulated quantization values instead of real quantization table.

### 3.1.3. Extraction module

The last part of the proposed network is the extraction module. The watermarked image will be fed into this network, and the network is responsible for extracting the watermark data. It is worth mentioning that the watermarked image may have been attacked before this step. The watermarked image will pass through consecutive convolutional layers without padding. Therefore, in each step, the size of the data will decrease until it achieves the same size as the main watermark size. The number of filters and the kernels' sizes are the same as the first part of the network of Fig.2. However, we do not use dilation in the extraction module. The extraction structure is illustrated in Fig. 4, which has a simpler form than the embedding module. Like the embedding structure, we have a reshape phase for images with a larger size than the network structure. After dividing the attacked image into the extraction network's acceptable size, a batch of its patches will be injected into the network. Finally, the reconstructed watermark will be produced as the output of the network.

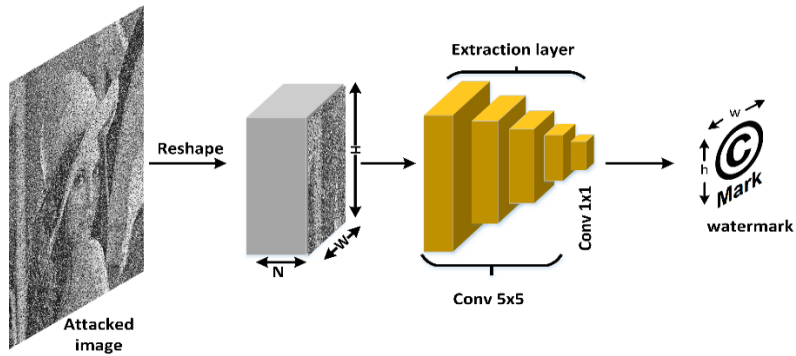

Fig. 4. The extraction module of the proposed method.

## 3.2. Network training and evaluating metrics

The whole structure is an end-to-end network that contains embedding and extraction modules. The encoder part consists of two networks: one with the cover image as input and the other one watermark data as input. The two networks work together. The extraction module accepts an attacked watermark image and extracts the hidden data. Initially, it seems that the two parts of the network (embedding and extraction network) work separately, and they also have the opposite aim. Since we have end-to-end network training and the attacks are a part of the network, the extraction's input is an attacked image. However, during the test phase, the attack layer will be removed and, we have a watermarked image as the input to this part. The embedding module is responsible for embedding the watermark data into the cover image with the lowest effect on the visual quality. High image quality requires weak embedding. On the other hand, the second part of the network should extract a watermark with the lowest difference compared to the original watermark. The extraction network is responsible for minimizing the bit error rate (BER) that is computed as (1):

$$BER(W, W') = \frac{\sum_{l=1}^{L_w} XOR(W(l), W'(l))}{L_W} \quad (1)$$

where $W$ and $W'$ are the original and extracted binary watermarks, respectively. Parameter $L_W$ indicates the length of the watermark. Based on Eq. 1, if the original and extracted watermarks are similar, the BER value should be near zero.

The similarity between the embedded and the extracted watermark requires a robust embedding that will not affect image quality. Since all parts work together, the loss function resolves the trade-off between transparency and robustness. It is a multi-objective network that the loss function is the combination of loss functions for the embedding and extraction modules:

$$L = \lambda_1 L_1 + \lambda_2 L_2 \quad (2)$$

where $\lambda_1$ and $\lambda_2$ are coefficients that control the trade-off between the two contrasting characteristics of the watermarked image. The image's visual quality and the extracted watermark act conversely. $L_1$ is the loss function that represents the difference between the cover image and the watermarked image. We used mean square error (MSE) as a comparing metric between the input image and the watermarked image:

$$L_1 = \frac{1}{n} \sum_{i=1}^{n} (img_i - img'_i)^2 \quad (3)$$

where $img$ is the cover image and $img'$ is the watermarked image. $L_1$ will be calculated for each training sample in a batch, and the average of them will be considered as the MSE value for a batch.

In Eq. 2, $L_2$ is the loss function which considers the difference between the input and the extracted watermark. We use binary-cross entropy and consider the extraction as a two-class classification that should label each pixel of the extracted watermark as zero or one. In Eq. 4, we compute the loss for each pixel, and it is performed for all extracted watermark bits.

$$L_2 = -\sum_{\substack{all \\ pixels}} t_i \log(p_i) + (1 - t_i)\log(1 - p_i) \quad (4)$$

where $p_i$ is the probability of the $i^{th}$ extracted watermark bit. $t_i$ is the known target value of the logo bit at the $i^{th}$ position. Hence $t_i$ is 0 or 1. The details of implementation and chosen parameters are in section 4.

## 4. EXPERIMENTAL RESULTS

The proposed method has been implemented on NVIDIA GeForce GTX 1080 Ti using Keras. We considered several experiments to show the performance of our method in terms of robustness and imperceptibility. The training phase is done using Pascal V.O.C. [49] and CIFAR10 [50]. We use the Granada dataset [51] that contains 49 grayscale images and the COCO dataset [52] for comparing our results with other methods. Our comparisons are made with well-known and concurrent works [44-46] to indicate its superiority.

The network structure and learning setup are explained in subsections 4.1 and 4.2. Subsections 4.3 reports the imperceptibility and robustness of the proposed method based on image processing metrics. We compare our approach with state-of-the-art techniques in subsection 4.4.

*4.1. Network configuration*

Based on Fig. 2, we have two networks in the embedding module. One network is for preparing the watermark data, and the second one is for embedding. The input image to WPN and embedding network have sizes $4 \times 4$ and $32 \times 32$, respectively. Hence, our cover image and watermark should be divided into blocks with these sizes and fed to networks as a batch of blocks. The WPN has four convolution layers. Each layer contains 64 filters with kernel size $2 \times 2$ and the same padding (each feature map has the same size as the input into the convolution layer), except the last layer that has one filter to concatenate to encoder output. Rectified Linear Unit (ReLU) is used as an activation function. The dilation convolution with size $2 \times 2$ is also used in each layer. The second network has the same network setting, but with four convolution layers, and in the last layer, we only have one filter and ReLU activation. The strides of all filters in embedding and extraction modules are set to one.

The decoder part in Fig. 2 has five convolution layers with the same setting as the encoder part. In the last layer, the filter size is $1 \times 1$ with one filter and a sigmoid activation function.

Each attack, such as salt and pepper, Gaussian noise, Gaussian blurring, and JPEG, are implemented as a layer, and the corresponding setting is performed based on the attacks. However, for implementing the JPEG attack, we have to divide the input image into $8 \times 8$ blocks. We implemented this using a convolution layer by 64 filters with size $8 \times 8$ and stride eight that these filters make the DCT basis in JPEG compression.

The extraction network has nine convolution layers. Each layer contains 64 filters with kernel size $5 \times 5$ without padding, except the last layer that has one filter and a sigmoid activation function to make the extracted watermark.

We use a $4 \times 4$ random binary string as the watermark data for embedding in a $32 \times 32$ block. However, based on the input image size, we replicate it to make the watermark the same size as the input image. If the input images had a size $512 \times 512$, we would need to divide them into $32 \times 32$ blocks. With this policy, the capacity will be $N \times 16$, where $N$ is the number of blocks. With this image size $N$=256, and the overall capacity for embedding is 4096 bits. We can divide the watermark data and embed it into several blocks. If we have a watermark string with a lower length, we can use redundancy in embedding and employ voting to extract the watermark. Consequently, the accuracy will improve.

*4.2. Learning setup*

We use a combination of CIFAR10 and Pascal V.O.C. for training. The first dataset has $32 \times 32$ color images, and the second one includes high-resolution color images. Hence, we randomly extract $32 \times 32$ patches with smooth to high-frequency patterns from each image of Pascal V.O.C. After converting to gray, the combination of CIFAR10 and Pascal V.O.C. has about 334K images. Ten percent of these images are used for the test phase. All images are normalized between [0,1] before training.

Kernel size and learning rate are hyper-parameters of a network that affect on network's performance. We can consider different scenarios for finding appropriate parameters. Here we first try to find the suitable kernel size. This helps us to have fewer searches by considering a specific optimizer. We use stochastic gradient descent (SGD) optimizer during training with the default learning rate ($lr = 0.01$) and change the kernel size. We also consider equal coefficients for each part of the loss function ($\lambda_1 =, \lambda_2 = 1$). With these settings, we try to find a suitable kernel size. Kernel size has an essential effect on network performance and parameter complexity. By increasing the kernel size, the number of learning parameters will be increased drastically. Table 1 shows the results of the proposed network with different kernel sizes. The PSNR and BER values are calculated as the average of five repetitions. Table 1 shows that the kernel sizes (2,2) and (5,5) have better results than others.

Table 1. Comparison of network performance with different kernel sizes.

|  | WPN | Embedding-Extraction network | WPN | Embedding-Extraction network | WPN | Embedding-Extraction network | WPN | Embedding-Extraction network | WPN | Embedding-Extraction network |
|---|---|---|---|---|---|---|---|---|---|---|
|  | (2,2) | (7,7) | **(2,2)** | **(5,5)** | (2,2) | (3,3) | (2,2) | (2,2) | (2,2) | (1,1) |
| PSNR | 33.37 | | **33.96** | | 28.33 | | 23.32 | | 23.86 | |
| BER | 0.000 | | **0.000** | | 0.000 | | 0.0 | | 0.399 | |
| Number of learning parameters | 1652300 | | 1171184 | | 296844 | | 184652 | | 64908 | |

SGD and Adam [53] are standard optimizers in deep network learning, but Adam usually has a higher speed in converging to global optimal [53]. To select the better optimizer, we trained the network with these optimizers and different learning rates. Then the accuracy of learned weights is evaluated on one of the standard watermark images [51]. The kernel size is selected based on Table 1 in this experiment. Table 2 shows the results of the network with different settings. When the learning rate is decreased, the network has better results on training data, but our result will not be acceptable on test data due to overfitting. If the learning rate is higher, then our network can not learn, and we will see a low performance. As we can observe, the network has better imperceptibility and robustness by Adam optimizer with the default value of learning rate (0.001).

Table 2. Comparison of network performance with optimizers and different learning rates.

| optimizer | | Learning rate | | | |
|---|---|---|---|---|---|
| | | 0.01 | **0.001** | 0.0001 | 0.00001 |
| **Adam** | BER | 0.004 | 0.000 | 0.000 | 0.000 |
| | PSNR | 32.87 | **38.06** | 37.96 | 35.21 |
| SGD | BER | 0.000 | 0.000 | 0.000 | 0.000 |
| | PSNR | 33.96 | 35.69 | 35.62 | 28.95 |

Loss function coefficients are another hyper-parameter in the network. Increasing each coefficient will improve the related output, with a negative effect on the other results. Hence, selecting suitable coefficients that resolve the trade-off between two outputs is vital. Since the image size is larger than the watermark, assigning a larger coefficient to the watermark image output is reasonable. Based on Table 1 and Table 2 the kernel size and optimizer are (2,2), (5,5), and Adam (lr=0.001), respectively, in this experiment. We calculate $\lambda_1$ and $\lambda_2$ experimentally and they are 0.95 and 0.05, respectively.

Table 3 shows the final settings which are used in our network. In the test phase, the end-to-end trained model is split into two parts without the attack layer, and the embedding and extraction networks are considered two distinct networks, as shown in Fig. 2 and Fig. 4. We examine the robustness of the framework by real attacks instead of simulated ones. We embed random 1024 bit (32 ×32) watermarks in 512 ×512 grayscale images to evaluate the trained networks. The cover image has a capacity of 4096 bit so that the watermark can be embedded with four times redundancy. Each watermark bit is repeated four times in a regular pattern in this plane to make a 64 ×64 bit redundant plane. The cover image and watermark plane are partitioned into 32 ×32 sub-images and 4 ×4 sub-watermarks for feeding to the embedding network based on network input size. The embedding network's output is watermarked sub-images that are tiled with each other to form the watermarked image. Then the watermarked image is passed through several attacks to evaluate the robustness of the proposed method. The extraction phase has the same procedure for preparing the attacked image. First, the attacked watermarked image is divided into 32 ×32 sub-images and is fed into the extraction network, which extracts 4 ×4 patches of the watermark plane. Afterward, these partitions are tiled to make a 64 ×64 redundant watermark plane. Eventually, a voting procedure is applied to the extracted watermark plane to produce the 1024-bit watermark data.

Table 3. network training parameters

| Description | Parameters | value |
|---|---|---|
| Training image patch size | (W,H) | (32,32) |
| Watermark size | (w,h) | (4,4) |
| Embedding-Extraction kernel size | (m,n) | (5,5) |
| WPN kernel size | (m,n) | (2,2) |
| Learning rate | lr | 0.001 |
| Optimizer | - | Adam |
| Iteration number | Itr | $10^6$ |
| Loss function weights | $\lambda_1, \lambda_2$ | 0.95,0.05 |

*4.3. Visual quality and robustness*

Figure. 5 shows examples of the watermarked images that have been produced with the proposed method. We used 1024-bit watermark data for embedding, and PSNR and SSIM values are indicated for each image. As we can see, the reported values are considerable and demonstrate that our network has successfully performed the embedding with the lowest artifacts.

To evaluate the robustness of the proposed framework, we consider several image processing attacks including Gaussian noise ($\sigma = \{5, 15, 25\}$), crop ($\{10,20,30\}$), median filter ($\{3,5\}$), sharpening ($\{1,5,10\}$),Gaussian blurring ($\{1,1.6,2\}$) salt-pepper ($\{2,6,10\}$), and grid-crop ($\{10,20,30\}$). The value of $\sigma$ in Gaussian noise indicates the standard deviation of the noise. The mentioned values in crop and salt-pepper attacks are the percentage of changed pixels. Grid crop is a new attack that randomly selects $8 \times 8$ blocks of image and changes them to zero [45]. It can evaluate the performance of the method for watermark scattering and data sharing. The mentioned scale in the resize attack resizes the image for resizing attack and then resizes back to the original size using bilinear interpolation. BER is used as an evaluation measure. To have more accurate results, we consider 20 random watermark strings and repeat our experiments for each watermark, and then the final output is the average value of all tests. For attacks such as Gaussian noise, salt-pepper, grid-crop, in which the results can be different each time, we calculate the average value for ten times repetition. To show the performance of our method, we report the BER values against different attacks for some standard images such as Barbara, Lena, Baboon, Living room, and Bridge. The average value of BER for all images in the Granada database [51] is also indicated in our reports. The watermark has 1024 bits in length, and the input images' size is 512×512. Based on the network structure, the image will be divided into 32×32 blocks, and finally, we have 256 blocks that each has 16 bits capacity. Hence, in total, we will have a 4096 bit capacity for embedding. We use this extra space for redundant embedding. Based on the input image size and the size of the watermark, the times of redundancy can be different. Regardless of the input image and watermark size, the extracted watermark can be achieved using voting. First, we aggregate the watermark string as (5):

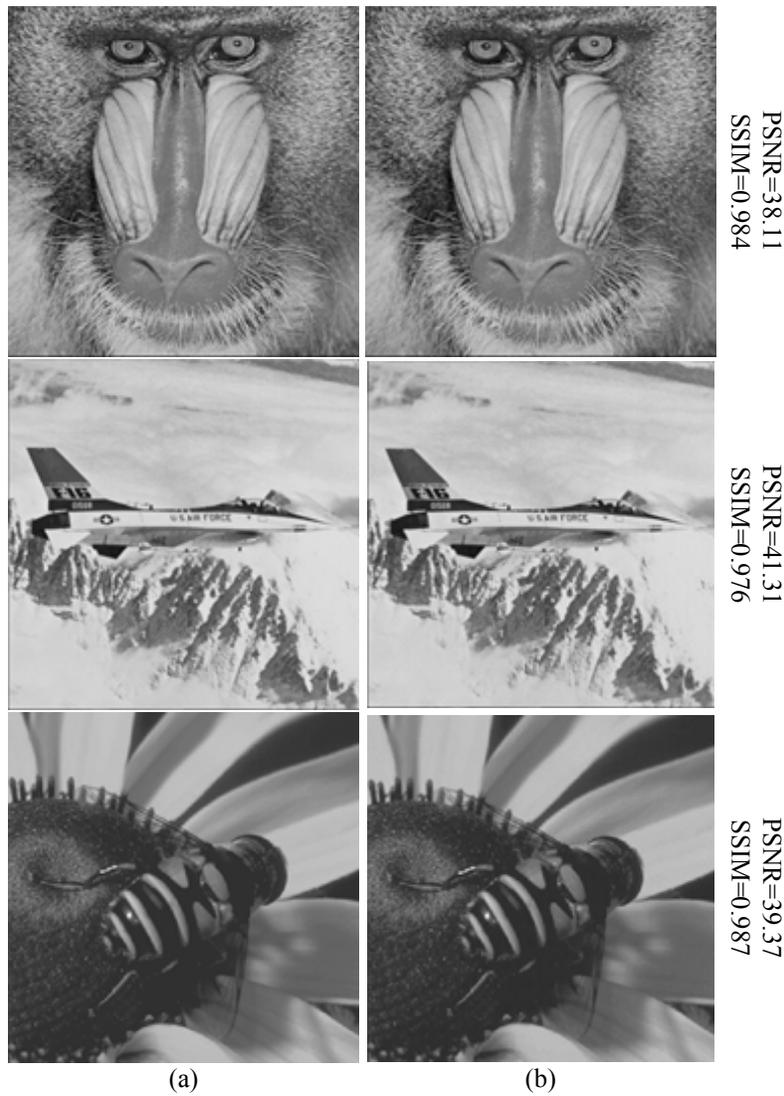

(a) (b)
Fig. 5. The visual quality of the watermarked image, (a) original image, (b) watermarked image.

$$V = \sum_{i=1}^{rn} W(i) \tag{5}$$

After aggregating to achieve the final watermark, we use voting as (6).

$$W' = \begin{cases} 1 & \text{if } V_i > rn/2 \\ 0 & \text{o.w} \end{cases} \tag{6}$$

where $rn$ is the number of redundancy. For input image of size $img_H \times img_W$, the number of blocks will be $N = (img_H \times img_W)/(32 \times 32)$, and the total capacity is $C = 16 \times N$. Assuming the length of a watermark is $l$, so $rn = C/l$. Our proposed method does not have a dependency on input image or watermark size, and it is enough that its size is a multiple of 32. Table 4 and Table 5 show the percentage of BER values against image processing attacks. Based on results, especially against crop and grid-crop attacks, we can conclude our network has a good ability for distributing the watermark in all parts of the cover image. By deleting random parts of the input image, the proposed method could extract a considerable percentage of the watermark data. Figure 6 shows image processing attacks on the Bridge image to show our method's performance. The BER values are also reported.

Table 4. Imperceptibility and robustness of proposed method based on BER, PNSR, and SSIM.

|  | Imperceptibility | | Robustness- BER (%) | | | | | | | | | | |
|---|---|---|---|---|---|---|---|---|---|---|---|---|---|
|  | PSNR | SSIM | Sharpening (radius) | | | JPEG | | | MED size | | Salt-pepper (%) | | |
|  |  |  | 1 | 5 | 10 | 90 | 70 | 50 | 3 | 5 | 2 | 6 | 10 |
| Barbara | 37.14 | 0.980 | 0.006 | 0.009 | 0.013 | 0.148 | 0.310 | 0.364 | 0.372 | 0.379 | 0.070 | 0.144 | 0.237 |
| Living room | 38.55 | 0.980 | 0.004 | 0.010 | 0.012 | 0.137 | 0.301 | 0.379 | 0.365 | 0.402 | 0.061 | 0.174 | 0.240 |
| Bridge | 36.77 | 0.985 | 0.004 | 0.026 | 0.040 | 0.056 | 0.252 | 0.338 | 0.386 | 0.412 | 0.058 | 0.132 | 0.215 |
| Baboon | 35.96 | 0.982 | 0.003 | 0.023 | 0.022 | 0.042 | 0.219 | 0.325 | 0.428 | 0.450 | 0.071 | 0.156 | 0.245 |
| Lena | 39.34 | 0.976 | 0.006 | 0.007 | 0.012 | 0.186 | 0.353 | 0.372 | 0.393 | 0.409 | 0.076 | 0.196 | 0.296 |
| Granada [51] | 40.34 | 0.991 | 0.001 | 0.008 | 0.014 | 0.062 | 0.293 | 0.379 | 0.373 | 0.491 | 0.048 | 0.185 | 0.341 |

Table 5. Imperceptibility and robustness of proposed method based on BER, PNSR, and SSIM.

|  | Imperceptibility | | Robustness- BER (%) | | | | | | | | | | | |
|---|---|---|---|---|---|---|---|---|---|---|---|---|---|---|
|  | PSNR | SSIM | Gaussian noise | | | Gaussian blurring (radius) | | | Crop (%) | | | Grid-crop (%) | | | No-attack |
|  |  |  | 5 | 15 | 25 | 1 | 1.6 | 2 | 10 | 20 | 30 | 10 | 20 | 30 |  |
| Barbara | 37.14 | 0.980 | 0.059 | 0.246 | 0.256 | 0.357 | 0.371 | 0.442 | 0.037 | 0.050 | 0.133 | 0.034 | 0.116 | 0.169 | 0.004 |
| Living room | 38.55 | 0.980 | 0.067 | 0.277 | 0.276 | 0.378 | 0.376 | 0.385 | 0.023 | 0.053 | 0.125 | 0.047 | 0.110 | 0.210 | 0.006 |
| Bridge | 36.77 | 0.984 | 0.045 | 0.216 | 0.220 | 0.350 | 0.387 | 0.386 | 0.018 | 0.079 | 0.103 | 0.018 | 0.097 | 0.177 | 0.005 |
| Baboon | 35.96 | 0.982 | 0.032 | 0.208 | 0.228 | 0.316 | 0.371 | 0.394 | 0.035 | 0.084 | 0.127 | 0.043 | 0.135 | 0.216 | 0.013 |
| Lena | 39.34 | 0.976 | 0.077 | 0.316 | 0.343 | 0.334 | 0.371 | 0.372 | 0.029 | 0.050 | 0.114 | 0.061 | 0.120 | 0.229 | 0.002 |
| Granada [51] | 40.34 | 0.991 | 0.051 | 0.327 | 0.405 | 0.281 | 0.416 | 0.473 | 0.025 | 0.066 | 0.110 | 0.001 | 0.004 | 0.005 | 0.002 |

*4.4. comparison with state-of-the-art*

Several new watermarking methods based on deep networks have been investigated to evaluate the proposed method's performance [44-46]. Each of these methods has used a different dataset and watermark length in their results; therefore, we use suitable watermark length based on their setting. The proposed method and comparable methods should have the same or near visual perceptuality to have fair comparisons.

The method presented in [44] is an end-to-end network that uses a GAN network to improve the reconstructed image quality. In this method, the watermark is a vector of length $L$. The input image also has dimensions of $M \times M$, and the watermark is repeated for each pixel of the image and is combined with the input image as a tensor with dimensions of $M \times M \times L$. In this case, each pixel of the image has the information of all watermark' bits, which reduces the image quality. They used the COCO database in their results, which has dimensions of $128 \times 128$, and the watermark's length is equal to 30. The images are color images in the YCbCr space. To compare our method with [44], we used similar cover images as [44] with a random watermark with a length of 32. After converting the cover image to YCbCr space, each of the color channels is input to our proposed network. Finally, by voting between extracted watermarks from each color channel, the final watermark is calculated.

The main idea in [45] is using the DCT layer as a preprocessing phase in their structure. They change the input image into a vector with the size MN×1. Then after vectorizing the watermark, they concatenate the watermark vector with the image vector. The final vector is fed into the DCT layer, and the DCT coefficients are used in their deep network. Adding the input image into the end of the encoder module as a residual helps their method to have a better imperceptibility. They also used a strength-factor in the test phase to control the strength of watermark embedding. Table 6 shows the proposed method's output against different attacks compared to other methods using the COCO database. SSIM and PSNR values of our approach are nearly the same as methods in Table 6 for a fair comparison. In a cropout attack, part of an image is randomly selected and replaced with part of the noise image. In a dropout attack, some pixels of the image are randomly selected and replaced with a pixel of noise image.

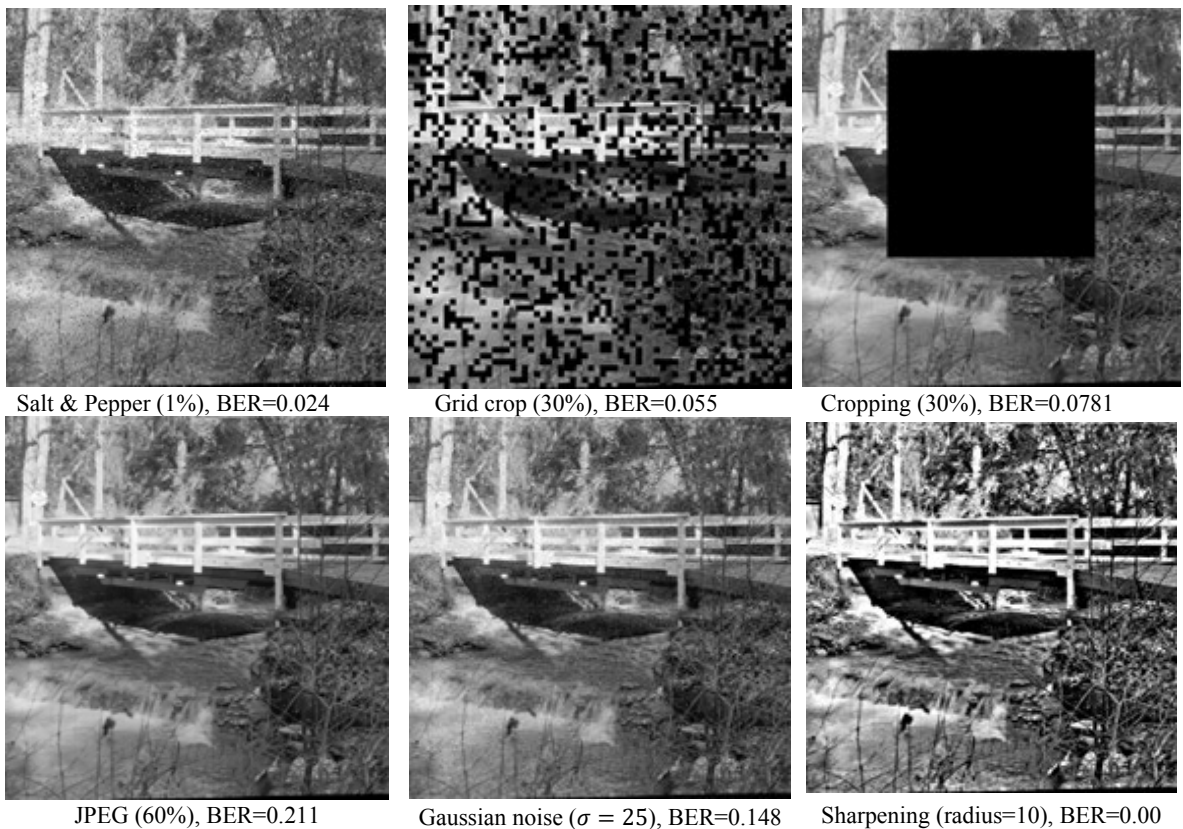

Salt & Pepper (1%), BER=0.024    Grid crop (30%), BER=0.055    Cropping (30%), BER=0.0781

JPEG (60%), BER=0.211    Gaussian noise ($\sigma = 25$), BER=0.148    Sharpening (radius=10), BER=0.00

Fig.6. The visual effect of various attacks on Bridge. The BER percent is shown for each image.

Table 6. Robustness of proposed method based on BER in [44-45] and the proposed method.

| Method | BER | | | | |
|---|---|---|---|---|---|
| | Gaussian Filter ($\sigma = 2$) | Crop (3.5%) | Dropout (30%) | Cropout (30%) | JPEG (50) |
| [44] | 0.040 | 0.120 | 0.070 | 0.060 | 0.370 |
| [45] | 0.500 | 0.000 | 0.080 | 0.075 | 0.254 |
| Proposed method | 0.490 | **0.000** | 0.093 | **0.017** | 0.263 |

As we can see, our method has better results against crop, cropout, and JPEG attack in comparison to [44]. Compared to [45] that uses accurate quantized coefficients in their method, our method has a considerable result against cropout attack and comparable results against other attacks. However, in Gaussian filter [44] has a better output, and it can be due to having a separate network for different attacks. Because our proposed network is multi-attack during learning, it can not have better results against all attacks.

Table 7 shows another comparison with [46] and [45] that have near PSNR (ours=40.34, [45]= 40.24, [46]=32.5) and SSIM (ours=0.991, [45]= 0.987, [46]=0.973). They used the Granada dataset. The watermark length is 1024 bits, and the input image size is 512×512. The method in [46] has the indicated value for PSNR 32.5, but our PSNR is 40.3, so our method with higher PSNR has considerable BER values for JPEG attack. Our method performs better in some attacks such as grid-crop that randomly some parts of images are deleted. This shows that the watermark is distributed inside the image, and by removing some parts, other parts of the image would still have the information. Our network is just trained on JPEG, Gaussian noise, and salt-pepper attacks, but based on Table 6. and Table 7. our method has considerable results against other attacks such as crop, dropout, cropout, and grid-crop. All these attacks try to remove part of the image and replace it with different data. This means part of the hidden information will remove. But due to expanding the watermark into all parts of the image, other parts of the image can help retrieve the watermark data by deleting some part of the watermark. Our method does not have better results against resizing in comparison to [45]. We do not consider resizing during training, and other used attacks also are not similar or near this attack. So our network is not very good in this attack.

Table 7. Robustness of proposed method based on BER in [45], [46] and the proposed method.

| Attack | | model | | |
|---|---|---|---|---|
| | | Ours | [45] | [46] |
| Gaussian noise (%) | 5 | **0.051** | 0.417 | - |
| | 15 | **0.327** | 0.456 | - |
| | 25 | **0.405** | 0.469 | - |
| Crop (%) | 10 | **0.025** | 0.077 | - |
| | 20 | **0.066** | 0.131 | - |
| | 30 | **0.111** | 0.188 | - |
| Salt-pepper (%) | 2 | **0.028** | 0.029 | - |
| | 6 | 0.185 | **0.045** | - |
| | 10 | 0.341 | **0.091** | - |
| Grid-crop (%) | 20 | **0.001** | 0.086 | - |
| | 30 | **0.004** | 0.130 | - |
| | 40 | **0.005** | 0.180 | 0.170 |
| JPEG (%) | 90 | **0.062** | 0.093 | 0.243 |
| | 70 | 0.293 | **0.164** | 0.289 |
| | 50 | 0.379 | **0.256** | - |
| Gaussian bluring (%) | 1 | 0.281 | **0.086** | - |
| | 1.6 | 0.416 | **0.390** | - |
| | 2 | **0.473** | 0.500 | - |
| MED (%) | 3 | 0.373 | **0.134** | - |
| | 5 | **0.491** | 0.500 | - |
| Sharpening (%) | 1 | **0.001** | 0.009 | - |
| | 5 | **0.008** | 0.024 | - |
| | 10 | **0.014** | 0.032 | - |
| Resize (%) | 0.5 | 0.390 | **0.212** | - |
| | 0.75 | 0.250 | **0.038** | - |
| | 1.5 | 0.048 | **0.026** | - |

## 5. CONCLUSION

In this paper, we proposed an end-to-end trained deep network for blind watermarking. This network contains two main modules: embedding and extraction. The embedding's responsibility is to make the input image ready for watermarking. The encoder part contains two networks that operate in parallel. One network accepts the input image, and after convolution layers, the high-level features of the image are extracted and ready for combining with a watermark. The second network accepts a watermark string and prepares the watermark based on the input image content. Finally, the watermark network's output is replicated to generate a matrix of the same size as the input image. The network's outputs are concatenated and are sent to the decoder network. Adding attack as a differentiable layer into the proposed structure helps the network to perform better against attacks and prevents the network from memorizing. Due to having a multi-object network with conflicting goals, the loss-function tries to resolve the trade-off between robustness and imperceptibility of the outputs. The proposed network is a multi-attack network. Multitasking means we consider a combination of different attacks. Multitasking helps the network to perform better against attacks during the test phase. Comparative results against the state-of-the-art methods demonstrate the superiority of the proposed method in robustness and visual perceptuality.

# REFERENCES

[1] A. Rashid, Digital watermarking applications and techniques: A brief review, *Int Journal of Computer Applications Technology and Research* 5(3) (2016), 147–150.

[2] W. Szepanski, "A signal theoretic method for creating forgery-proof documents for automatic verification," *In Proc. of Carnahan Conference on Crime Countermeasures*, pp. 101-109, 1979.

[3] J. Liu and X. He, "A review study on digital watermarking," *in Proc IEEE International Conference on Information and Communication Technology*, pp. 337–34, 2005.

[4] A. Shehab, M. Elhoseny, K. Muhammad, A. K. Sangaiah, P. Yang, H. Huang, G. Hou, "Secure and robust fragile watermarking scheme for medical images," *IEEE Access*, pp. 10269–10278, 2018.

[5] Y. Cheng, "Music database retrieval based on spectral similarity," *2$^{nd}$ Int. Symposium on Music Information Retrieval (IS-MIR)*, 2001.

[6] M. Faundez-Zanuy, M. Hagmüller, G. Kubin, "Speaker identification security improvement by means of speech watermarking," *Pattern Recognition*, pp. 3027–3034, 2007

[7] R. S. Broughton, W. C. Laumeister, "Interactive video method and apparatus," 1989.

[8] V.M. Potdar, S. Han and E. Chang, "A survey of digital image watermarking techniques," *in Proc IEEE International Conference on Industrial Informatics*, pp. 709–716, 2005.

[9] D.G. Savakar, A. Ghuli, "Non-blind digital watermarking with enhanced image embedding capacity using Dmeyer wavelet decomposition , SVD and DFT," *pattern Recognition and Image Analysis,* pp. 511-517, 2017.

[10] G. Anbarjafari, C. Ozcinar, "Imperceptible non-blind watermarking and robustness against tone mapping operation attacks for high dynamic range images," *Multimedia Tools and Applications*, pp. 24521–24535, 2018.

[11] R.-Z. Wang, C.-F. Lin, J.-C. Lin, "Image hiding by optimal LSB substitution and genetic algorithm," *Pattern Recognition*, pp.671–683, 2001.

[12] T. Zong, Y. Xiang, I. Natgunanathan, S. Guo, W. Zhou, and G. Beliakov, "Robust histogram shape-based method for image watermarking," *IEEE Transactions on Circuits and Systems for Video Technology,* pp. 717–729, 2014.

[13] H. Zhang, H. Shu, G. Coatrieux, J. Zhu, Q.J.Wu, Y. Zhang, H. Zhu, and L. Luo, "Affine Legendre moment invariants for image watermarking robust to geometric distortions," *IEEE Transactions on Image Processing*, pp. 2189–2199, 2011.

[14] M. Tagliasacchi, G. Valenzise and S. Tubaro, "Hash-based identification of sparse image tampering," *IEEE Trans Image Process*, pp. 2491–2504, 2009.

[15] V. Solachidis and P. Ioannis, "Circularly symmetric watermark embedding in 2-D DFT domain," *IEEE Transactions on* Image *Processing*, pp. 1741–1753, 2001.

[16] A.H. Taherinia and M. Jamzad, "A robust spread spectrum watermarking method using two levels DCT," *International Journal of Electronic Security and Digital Forensics*, pp. 280–305, 2009.

[17] SA Parah, J.A. Sheikh, N.A. Loan and G.M. Bhat, "Robust and blind watermarking technique in DCT domain using inter-block coefficient differencing," *Digital Signal Processing*, pp. 11–24, 2016.

[18] SJ Horng, D. Rosiyadi, P. Fan, X. Wang and M.K. Khan, "An adaptive watermarking scheme for e-government document images," *Multimedia Tools and Applications*, pp. 3085–3103, 2014.

[19] C. Tian, R.H. Wen, W.P. Zou and L.H. Gong, "Robust and blind watermarking algorithm based on DCT and SVD in the contourlet domain," Multimedia Tools and Applications, pp.1-27, 2019.

[20] B. Jagadeesh, P. Rajesh Kumar, and P. Chenna Reddy, "Robust digital image watermarking based on fuzzy inference system and backpropagation neural networks using DCT," *Soft Computing*, pp. 3679–3686, 2016.

[21] B. Jagadeesh, P.R. Kumar, and P.C. Reddy, "Fuzzy inference system based robust digital image watermarking technique using discrete cosine transform," *Procedia Computer Science*, pp. 1618–1625, 2015.

[22] V.M. Potdar, S. Han and E. Chang, "A survey of digital image watermarking techniques," *in Proc IEEE International Conference on Industrial* Informatics, pp. 709– 716, 2005.

[23] P. Rasti, G. Anbarjafari, and H. Demirel, "Colour image watermarking based on wavelet and QR decomposition," *In Proc of IEEE Signal Processing and Communications Applications Conference*, pp. 1–4, 2017.

[24] E. Nezhadarya, Z.J. Wang and R.K. Ward, "Robust image watermarking based on multi-scale gradient direction quantization," *IEEE Transactions on Information Forensics and Security*, pp. 1200–1213, 2011.

[25] M. Hamghalam, S. Mirzakuchaki and M.A. Akhaee, "Geometric modeling of the wavelet coefficients for image watermarking using optimum detector," *IET Image Processing*, pp. 162–172, 2014.

[26] M.Jamali, S. Rafiei, S.M. Soroushmehr, N. Karimi, S. Shirani, K. Najarian and S. Samavi, "Adaptive image watermarking using human perception based fuzzy inference system. Journal of Intelligent & Fuzzy Systems," *Journal of Intelligent & Fuzzy Systems*, pp.4589-4608, 2018.

[27] N. M. Makbol, B. E. Khoo, T. H. Rassem, K. Loukhaoukha, "A new reliable optimized image watermarking scheme based on the integer wavelet transform and singular value decomposition for copyright protection," *In Formation Sciences,* pp. 381–400, 2017.

[28] H.R. Fazlali, S. Samavi, N. Karimi and S. Shirani, "Adaptive blind image watermarking using edge pixel concentration," *Multimedia Tools and Applications*, pp. 3105–3120, 2017.

[29] S. Etemad, M. Amirmazlaghani, "A new multiplicative watermark detector in the contourlet domain using t location-scale distribution," *Pattern Recognition,* pp. 99–112, 2018.

[30] A. Akhaee, S.M. Sahraeian, and F. Marvasti, "Contourlet based image watermarking using optimum detector in noisy environment," *IEEE Transactions on Image Processing,* pp. 967–980, 2010.

[31] H. Sadreazami, M.O. Ahmad and M.N. Swamy, "Multiplicative watermark decoder in contourlet domain using the normal inverse Gaussian distribution," *IEEE Transactions on Multimedia*, pp. 196–207, 2016.

[32] J. Li, C. Yu, B. B. Gupta, X. Ren, "Color image watermarking scheme based on quaternion Hadamard transform and Schur decomposition," *Multimedia Tools and Applications*, pp. 4545–4561, 2017.

[33] E. Etemad, S. Samavi, S.R. Soroushmehr, N. Karimi, M. Etemad, S. Shirani, and K. Najarian, "Robust image watermarking scheme using bit-plane of Hadamard coefficients," *Multimedia Tools and Applications*, pp. 1–23, 2017.

[34] M. Heidari, S. Samavi, S. M. R. Soroushmehr, S. Shirani, N. Karimi, K. Najarian, "Framework for robust blind image watermarking based on classification of attacks," *Multimedia Tools and Applications*, pp. 23459–23479, 2017.

[35] R. P. Singh, N. Dabas, V. Chaudhary, "Online sequential extreme learning machine for watermarking in DWT domain," *Neurocomputing*, pp. 238–249, 2016.
[36] A. M. Abdelhakim, M. Abdelhakim, "A time-efficient optimization for robust image watermarking using machine learning," *Expert Systems with Applications*, pp. 197–210, 2018.
[37] A. Khan, S. F. Tahir, A. Majid, T.-S. Choi, "Machine learning based adaptive watermark decoding in view of anticipated attack," *Pattern Recognition*, pp. 2594–2610, 2008.
[38] S. Ren, K. He, R. Girshick, J. Sun, "Faster R-CNN: Towards real-time object detection with region proposal networks," *in Advances in neural information processing systems*, pp. 91–99, 2015.
[39] G. Huang, Z. Liu, L. Van Der Maaten, K. Q. Weinberger, "Densely connected convolutional networks," *In Proceedings of the IEEE conference on computer vision and pattern recognition*, pp. 4700-4708, 2017.
[40] Z. Zheng, L. Zheng, Y. Yang, "A discriminatively learned CNN embedding for person re-identification," *ACM Transactions on Multimedia Computing, Communications, and Applications (TOMM)*, pp. 1-20, 2017.
[41] C. Jin and S. Wang, S, "Applications of a neural network to estimate watermark embedding strength*," In Eighth International Workshop on Image Analysis for Multimedia Interactive Services (WIAMIS'07)*, pp. 68-68, 2007.
[42] H. Kandi, D. Mishra, S. R. S. Gorthi, "Exploring the learning capabilities of convolutional neural networks for robust image watermarking," *Computers & Security*, pp. 247–268, 2017.
[43] S.M. Mun, S.H. Nam, H.U. Jang, D. Kim, H.K. Lee, "A robust blind watermarking using convolutional neural network," arXiv preprint arXiv:1704.03248, 2017.
[44] J. Zhu, R. Kaplan, J. Johnson, L. Fei-Fei, "Hidden: Hiding data with deep networks," *In Proceedings of the European conference on computer vision (ECCV),* pp. 657-672, 2018.
[45] M. Ahmadi, A. Norouzi, S.M. Soroushmehr, N. Karimi, K. Najarian, S. Samavi and A. Emami, "ReDMark: Framework for residual diffusion watermarking based on deep networks," *Expert Systems with Applications*, pp. 113157, 2020.
[46] G. Hua, L. Zhao, H. Zhang, G. Bi, and Y. Xiang, "Random matching pursuit for image watermarking*," IEEE Transactions on Circuits and Systems for Video Technology*, pp.625-639, 2018.
[47] S. Ioffe, and C. Szegedy, "Batch normalization: Accelerating deep network training by reducing internal covariate shift," *In International conference on machine learning*, pp. 448-456, 2015.
[48] F. Yu, and V. Koltun, "Multi-scale context aggregation by dilated convolutions," arXiv preprint arXiv:1511.07122, 2015
[49] M. Everingham, S.A. Eslami, L. Van Gool, C.K. Williams, J. Winn, and A. Zisserman, "The pascal visual object classes challenge: A retrospective," *International journal of computer vision*, pp.98-136, 2015.
[50] A. Krizhevsky, V. Nair, G. Hinton, The CIFAR-10 dataset, online: http://www. cs. toronto. edu/kriz/cifar. Html
[51] Dataset of standard 512512 grayscale test images. URL http://decsai.ugr.es/cvg/CG/base.htm.
[52] T.Y. Lin, M. Maire, S. Belongie, J. Hays, P. Perona, D. Ramanan, P. Dollár, and C.L. Zitnick, "Microsoft coco: Common objects in context," *In European conference on computer vision,* pp. 740-755, 2014.
[53] DP Kingma and J. Ba, "Adam: A method for stochastic optimization," arXiv preprint arXiv:1412.6980, 2014.